\documentclass[a4paper, 12pt]{article} % Font size (can be 10pt, 11pt or 12pt) and paper size (remove a4paper for US letter paper)
\usepackage[right=1.1in,left=1.1in,top=1.1in,bottom=1.1in]{geometry}
\usepackage{graphicx} 
\usepackage{braket}
\usepackage{mathrsfs}
\usepackage[T1]{fontenc} % Required for accented characters
\usepackage{amsmath}
\usepackage{pxfonts}
\usepackage[T1]{fontenc}
\usepackage{amssymb}
\usepackage{natbib}
\usepackage{epstopdf}
\usepackage{setspace}
\usepackage{enumitem}
\usepackage{sectsty}
\usepackage{wrapfig} % Allows in-line images
\sectionfont{\large}
\bibpunct{(}{)}{,}{a}{}{,}
\usepackage{tikz}   %TikZ is required for this to work.  Make sure this exists before the next line
\usepackage{tikz-3dplot} %requires 3dplot.sty to be in same directory, or in your LaTeX installation
\usepackage[protrusion=true,expansion=true]{microtype} % Better typography

\newcommand{\qed}{\nobreak \ifvmode \relax \else
      \ifdim\lastskip<1.5em \hskip-\lastskip
      \hskip1.5em plus0em minus0.5em \fi \nobreak
      \vrule height0.75em width0.5em depth0.25em\fi}

\usepackage{bbm}
\usepackage{MnSymbol}
\usepackage[all]{xy}

\usepackage{xcolor,colortbl,array,amssymb}

\makeatletter
\title{\textbf{The Cosmic Void}} % Subtitle

\author{Eddy Keming Chen\thanks{Department of Philosophy,  University of California, San Diego, 9500 Gilman Dr, La Jolla, CA 92093-0119. Website: www.eddykemingchen.net. Email: eddykemingchen@ucsd.edu  }}

\date{Sara Bernstein and Tyron Goldschmidt (eds.), \textit{Non-Being: New Essays on the Metaphysics of Nonexistence},   Oxford University Press, 2021}

\begin{document}
\bibliographystyle{authordate1}

\maketitle % Print the title section

%----------------------------------------------------------------------------------------
%	ABSTRACT AND KEYWORDS
%----------------------------------------------------------------------------------------

\begin{abstract}
What exists at the fundamental level of reality? On the standard picture, the fundamental reality contains (among other things) fundamental matter, such as particles, fields, or even the quantum state. Non-fundamental facts are  explained by facts about fundamental matter, at least in part. In this paper, I introduce a non-standard picture called the ``cosmic void'' in which the universe is devoid of any fundamental material ontology. Facts about tables and chairs are recovered from a special kind of laws that satisfy  \emph{strong determinism}. All non-fundamental facts are  completely explained by  nomic facts. I discuss a concrete example of this picture in a strongly deterministic version of the many-worlds theory of quantum mechanics. I discuss some philosophical and scientific challenges to this view, as well as some connections to ontological nihilism.   

\end{abstract}

\hspace*{3,6mm}\textit{Keywords: ontological nihilism, objects, emergence, many-worlds, time's arrow, Humeanism, non-Humeanism, laws of nature, fundamentality, the Wentaculus}   % Keywords

%\begingroup
%\singlespacing
%\tableofcontents
%\endgroup

%\vspace{30pt} % Some vertical space between the abstract and first section

%----------------------------------------------------------------------------------------
%	ESSAY BODY
%----------------------------------------------------------------------------------------

%------------------------------------------------
\nocite{albert2000time,  }

\section{Introduction}

One of the hardest questions in fundamental physics and fundamental metaphysics is this: what exists at the fundamental level of reality? There is no consensus in contemporary physics. There is  much less  consensus in metaphysics.  

Nevertheless, on the physics side, we usually assume that the fundamental level of reality will be something physical and material. We assume that the fundamental level will not consist in purely mental ideas. We also assume that the fundamental level will not be completely devoid of matter, since our world is manifestly not empty. Different physical theories postulate different kinds of fundamental matter (and sometimes different versions of the same theory will differ in the fundamental material ontologies). To see some examples, here is a preliminary list of fundamental material ontologies in different physical theories: 
\begin{itemize}
	\item Newtonian mechanics: point particles;
	\item Maxwellian electrodynamics: charged particles and electromagnetic fields;
	\item General relativity: matter field on space-time;
	\item Quantum mechanics: a quantum state and / or local ontologies such as matter density field, flashes, or particles;
	\item Quantum field theory: a quantum state and / or local ontologies such as particles, particle number ontology, or fields;
	\item Loop quantum gravity: spin-foams;
	\item String theory: one-dimensional strings.  
\end{itemize}
(This list is certainly incomplete: future theories may have radically different ontologies; and even  theories on this list can be compatible with other ontologies.)

Describing the fundamental material ontology is an important task in theoretical physics, since fundamental matter plays crucial roles in a physical theory. First, the fundamental material ontology in a theory plays an explanatory and metaphysical role. It is the ultimate explanation for non-fundamental ontologies and non-fundamental facts. We use fundamental matter as the ultimate reduction base: we try to reduce the behaviors of macroscopic systems such as tables and chairs to the behaviors of microscopic constituent objects such as particles and fields. How tables and chairs move around can be explained by how the constituent particles and fields behave. How particles and fields behave can be derived from the laws of physics. If the particles and fields are part of the theory's fundamental material ontology, then the reduction is complete. If they are further reducible to something else in the fundamental material ontology, such as the quantum state, then more reduction is in order. In any case, it is hard to see how to carry out the reduction without the fundamental material ontology. 

Second, the fundamental material ontology plays a semantic role by being the subject matter of the laws of nature.  The terms in the laws of nature refer to properties of the fundamental material ontology. For example, the mass term in $F=ma$ refers to a property of Newtonian point particles. The acceleration term describes the change of their velocities. It is hard to see how fundamental physical laws make sense without the fundamental material ontology. 

Third, the fundamental material ontology plays an informational role in the physical theory. The physical laws are supposed to be simple. But how can such simple laws explain such a wide range of complicated phenomena? How does so much information come from such simple laws? It is made possible by appealing to the complicated initial conditions and boundary conditions of physical systems. The complicated data is  stored not in the laws but in the matter distribution, such as the locations of particles and configurations of fields. By postulating a fundamental material ontology, complicated phenomena can be derived from simple laws plus certain contingent (and complicated) facts about fundamental matter. There is a tight connection between fundamental material ontology and laws of physics that we will try to make more precise later. But it is worth noting that we usually put the ``mess'' in the fundamental material ontology so that the laws can be simple.

Those roles are important and seemingly irreplaceable. Hence, the fundamental material ontology seems indispensable to any successful physical theory.\footnote{Quantum theory presents an apparent counterexample, as the followers of the ``Copenhagen interpretation'' deny the existence of a determinate microscopic ontology. But we have made sufficient progress about the quantum measurement problem to understand that there are better ways to understand quantum theory. See \S3.1.  } Given those roles fundamental matter seems to play in successful physical theories, we also have reasons to postulate them in our fundamental metaphysics.

 However, it can be intellectually healthy to reexamine our assumptions. As \cite{TurnerON}  puts it, ``we can come to better understand a proposition by studying its opposite.'' We ask the following question: is a fundamental material ontology really indispensable to any successful physical theory? Reflecting on this question does not require us to reject scientific realism. It may even help us better appreciate the place of fundamental material ontology in our physical theories and metaphysical frameworks. 
 
In this paper, I discuss a possibility that I call ``the cosmic void.'' In the cosmic void scenario, the universe is devoid of any fundamental matter.\footnote{Hence, it is different from the notion of the cosmic void in astrophysics that the large-scale structure of the universe can be characterized by mostly empty space between galaxies.} I focus on the informational role of the fundamental matter ontology and suggest that, in certain theories, it can be played by the laws of physics.   To recover the non-fundamental facts---the manifest image of tables, chairs, and computers---I make use of \emph{strongly deterministic} laws of nature that determine a unique history of the universe. In this scenario, all of the non-fundamental facts are ultimately explained by nomic facts. I discuss a concrete example of the cosmic void scenario that arises when we try to put together a many-worlds theory of quantum mechanics in a time-asymmetric universe. The physical theory turns out to be strongly deterministic. To be sure, there are both philosophical and scientific challenges to such a possibility, especially regarding the metaphysical role and the semantic role of the fundamental matter ontology. I introduce the cosmic void scenario \emph{not to endorse it but to draw our attention to an interesting area of logical space that deserves more scrutiny.}

There are some similarities between this project and the anti-object metaphysics of ontological nihilism \citep{HawthorneCortensTON, TurnerON}, structural realism \citep{ladyman2007every}, generalism \citep{DasguptaI}, and the bare facts framework \citep{Maxwell2019}. However, the cosmic void possibility is at the same time more radical and more modest. It is more radical in that we do not postulate even fundamental properties, general facts, bare facts, or structural facts in the fundamental ontology. It is more modest in that the strategy is not supposed to work in general, but only in some special physical theories where strong determinism holds. Hence, the cosmic void scenario has a more restricted scope of application.

\section{The Possibility of the Cosmic Void}

What is the cosmic void? It is the scenario in which \emph{nothing material exists at the fundamental level}. Suppose spacetime is fundamental. Then, in this scenario, fundamentally speaking the spacetime is completely empty. There are no fundamental particles, fundamental fields, fundamental quantum states,  any kind of fundamental material distributions, or any non-trivial decorations on spacetime. 

This cosmic void scenario is allowed by both special and general relativity. However, that is not the intended interpretation here. In the empty solutions allowed by special and general relativity, non-fundamental facts that there are tables and chairs are false. In the cosmic void scenario I would like to describe, those non-fundamental facts remain \emph{true}, even though there is no fundamental material ontology. 

\begin{figure}
\centerline{\includegraphics[scale=0.28]{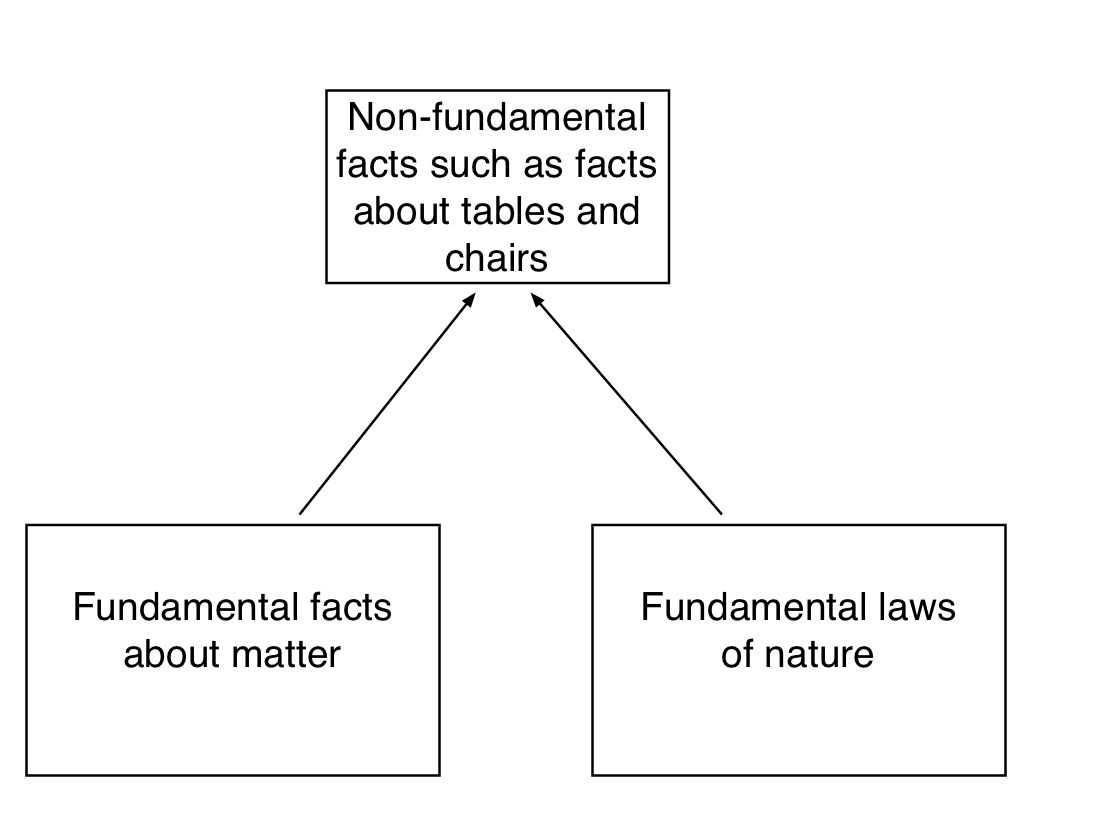}}
\caption{The standard picture where non-fundamental facts are explained by fundamental facts about matter and laws of nature. }
\end{figure}

\begin{figure}
\centerline{\includegraphics[scale=0.28]{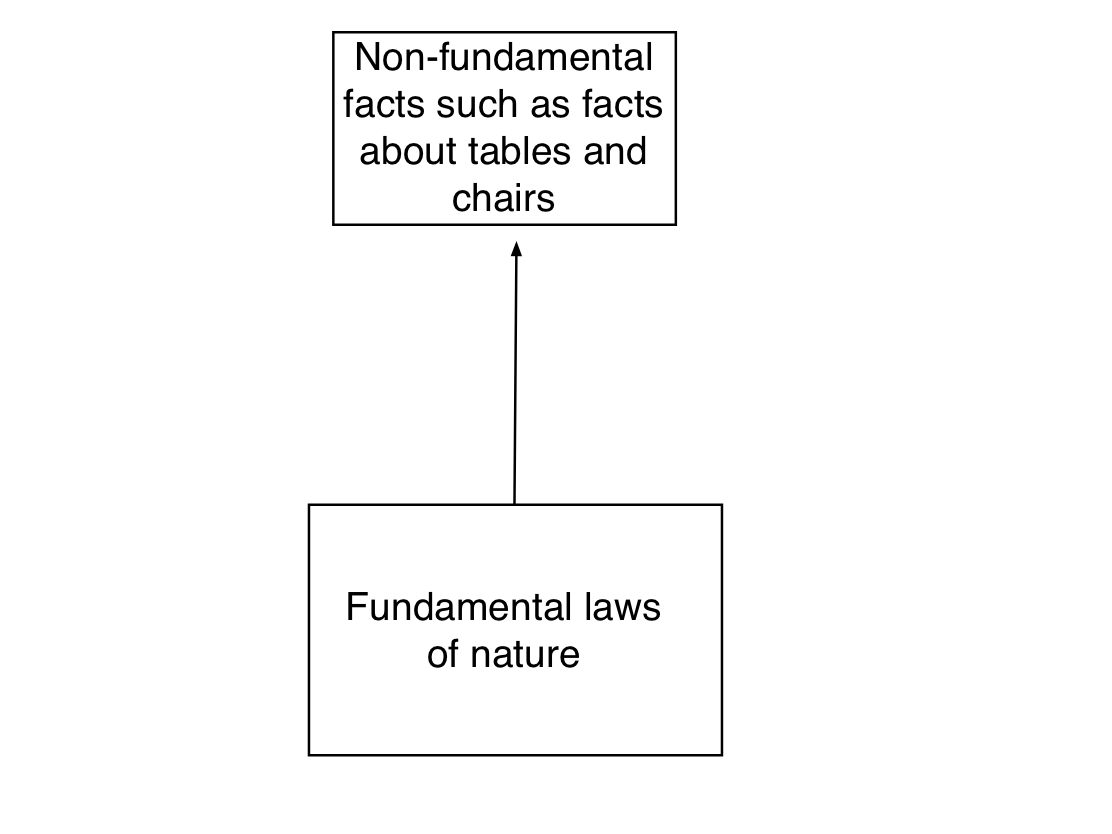}}
\caption{The non-standard picture where non-fundamental facts are completely explained by fundamental laws of nature. }
\end{figure}

On the standard picture (see Figure 1), non-fundamental facts about tables and chairs are made true by fundamental laws of nature and fundamental facts about the existence and behavior of fundamental matter, such as point particles. Fundamental particles constitute tables and chairs, and the existence and behavior of tables and chairs can be explained by the existence and behavior of the fundamental material ontology, together with the laws. 

On the non-standard picture (see Figure 2), since there is nothing material at the fundamental level, we have to appeal to something else to explain the non-fundamental facts. What can that explanation be? A candidate is the fundamental laws of nature. For the purpose of this paper, we  commit to a non-Humean theory of lawhood. Here we assume that laws are not merely summaries of the mosaic. There can be non-trivial laws even when the mosaic is completely empty and un-decorated. 

Usually, fundamental laws of nature are \emph{at best} partial explanations for most non-fundamental facts. The existence of a table in front of me at this point is not entailed by the standard laws of physics. At the very least, we need also the complete state of the universe at some time (or during some time interval). The complete state of the universe is not encoded in the dynamical laws, for they are compatible with many different states of the universe. In standard physical theories, the complete state of the universe refers to the state of the fundamental material ontology. Hence, the non-standard picture is not accommodated by the usual laws of physics.

\begin{figure}
\centerline{\includegraphics[scale=0.25]{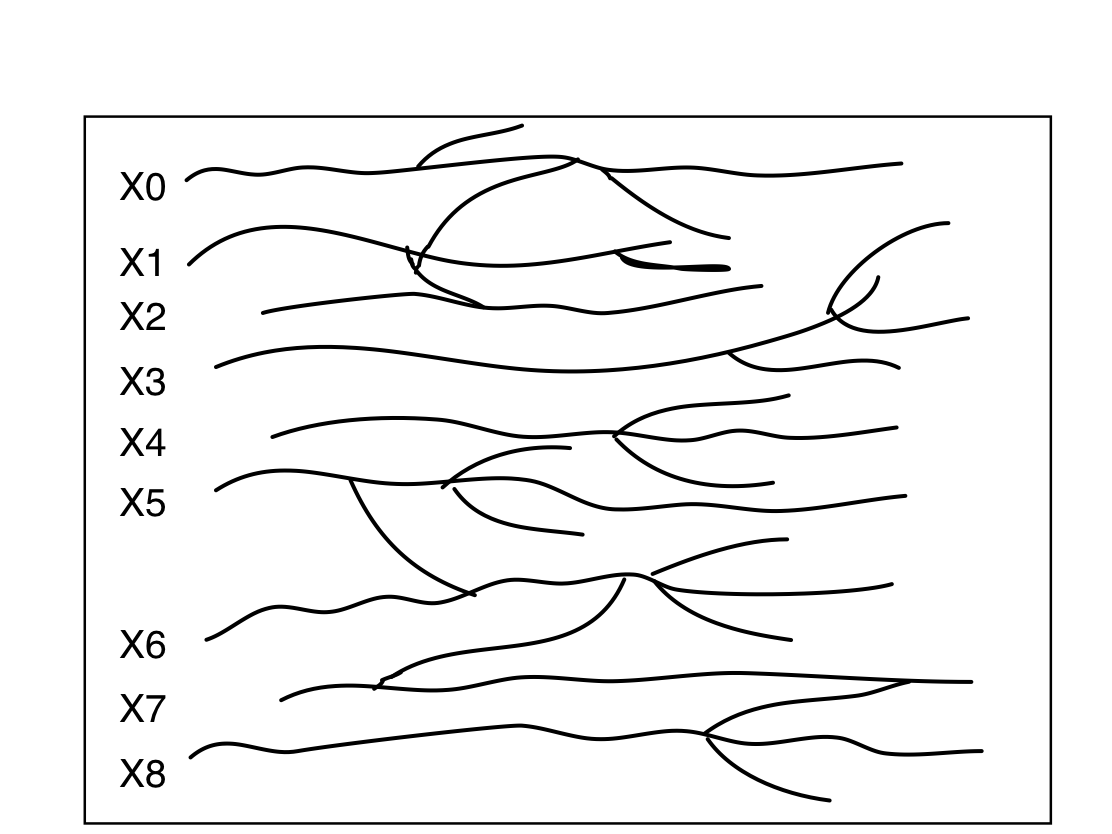}}
\caption{Indeterministic laws of physics. $X0$--$X8$ refer to different initial conditions of the universe. (Here the illustration is schematic. Usually there are infinitely many possible initial conditions.) The horizontal curves correspond to different histories of the universe. Different histories of the universe can overlap at some time and diverge later. Fixing an initial condition of the universe does not fix a unique history of the universe.}
\end{figure}

\begin{figure}
\centerline{\includegraphics[scale=0.25]{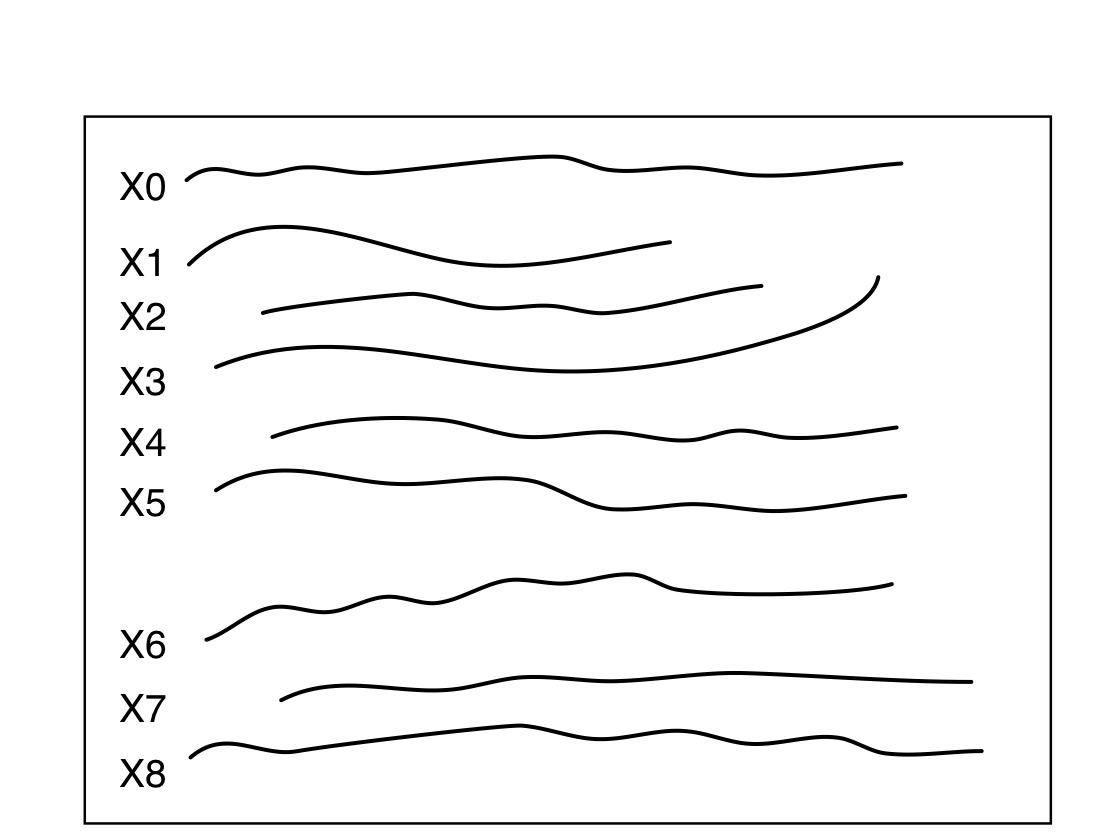}}
\caption{Deterministic laws of physics. $X0$--$X8$ refer to different initial conditions of the universe. The horizontal curves correspond to different histories of the universe. Different histories of the universe cannot overlap at any point in time. Fixing an initial condition of the universe fixes a unique history of the universe.}
\end{figure}

It is useful to distinguish between two  types of laws of physics: indeterministic laws and deterministic laws. If the laws are indeterministic (see Figure 3), then given a complete state of the universe at some time (or some  duration of time), the laws can allow multiple different pasts and futures of the universe. In other words, different histories can overlap. If the laws also assign objective probabilities or chances to the histories, then given a complete state of the universe at some time, the laws assign a unique probability distribution over future (or past) histories. 

If the laws are deterministic (see Figure 4), then given a complete state of the universe at some time (or some  duration of time), the laws allow only one past and one future of the universe. In other words, different histories cannot overlap: no two distinct histories of the universe can overlap (be exactly the same in terms of the distribution of microphysical properties and ontologies) at any point in time. Hence, the deterministic theory can be more informative than an indeterministic theory in the sense that the former but not the latter uniquely determines the entire history of the universe given the dynamical laws and a specification of the material ontology at some time (or some  duration of time). In a universe governed by deterministic laws of physics, fixing the initial condition of the universe is sufficient to fix the entire history of the universe. 

\begin{figure}
\centerline{\includegraphics[scale=0.25]{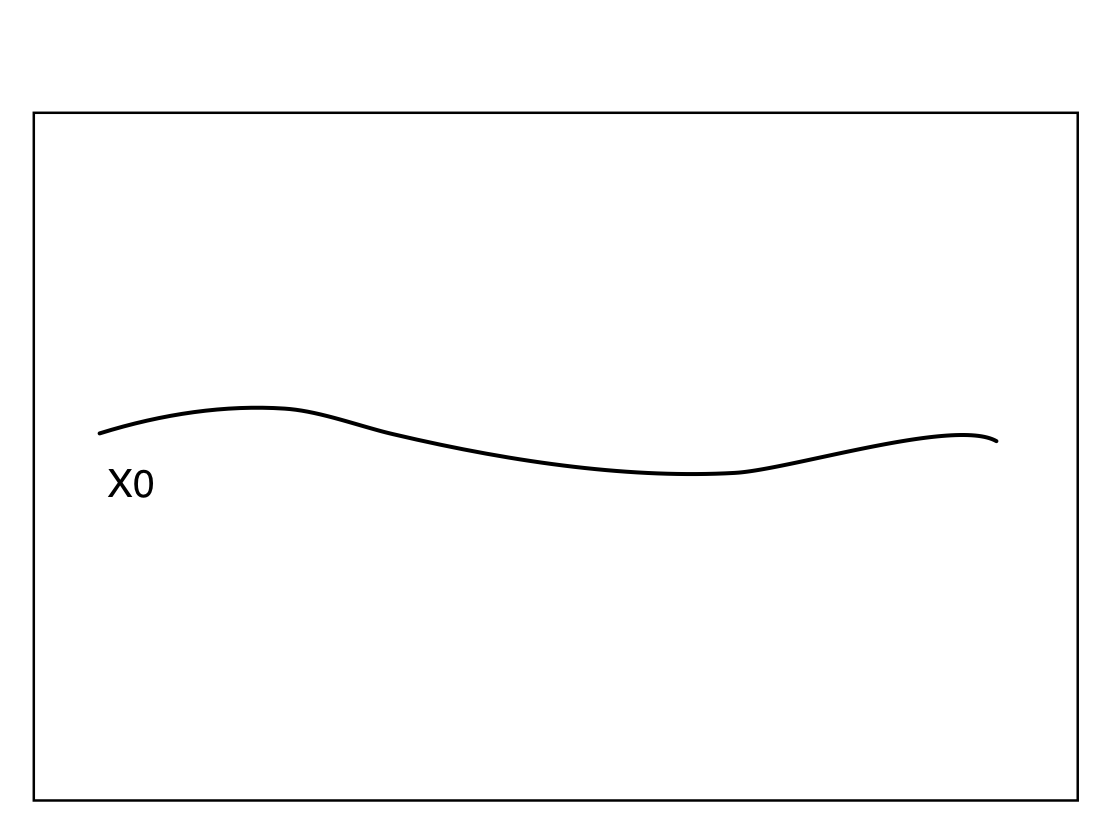}}
\caption{Strongly deterministic laws of physics. $X_0$ refers to the unique initial condition allowed by the laws. The horizontal curve refers to the unique history of the universe allowed by the laws. The laws completely fix the history of the universe.}
\end{figure}

However, information about the exact initial condition itself is not contained in any standard law of physics. Usually it is  the distribution of the fundamental material ontology, not the laws, that specify  the complete initial condition. After all, usually the exact initial condition is too complicated to express by any simple law. (As we discuss below, this is true even after adding the Past Hypothesis as an additional fundamental law of nature, for it only pins down the macroscopic initial condition of the universe and not the microscopic initial condition.) Hence, it is in this sense that the standard deterministic laws of physics are compatible with many different possible worlds.  Usually the laws do not pick out a unique world. 

The situation is transformed when the laws  are \emph{strongly deterministic} in the following sense: 

\begin{description}
\item[Strong Determinism] The laws of nature pick out a unique history of the universe. 
\end{description}
A world is strongly deterministic if its fundamental laws pick out a unique history of the universe (see Figure 5).\footnote{This notion of strong determinism is introduced in \citep{roger1989emperor}. This is different from the notion of ``superdeterminism'' that is sometimes invoked in the context of avoiding Bell non-locality. See \citep{ChenBell} for an overview. }  One way the laws can be strongly deterministic  is by satisfying two conditions: 
\begin{enumerate}
	\item The laws are deterministic.
	\item The laws pick out a unique initial condition. 
\end{enumerate}
Here, the unique initial condition refers to an exact specification of the microphysical details of the universe at $t_0$. Typically, it is difficult to pick out a unique (microscopic) initial condition without making the laws overly complicated (so that they no longer qualify as good candidate fundamental laws of physics). Hence, it is highly non-trivial to write down a \emph{simple} law (or set of laws) that can do that. In \S3, we look at a concrete example of a law---the Initial Projection Hypothesis---that is not only simple but also selects a unique (microscopic) initial condition in the Everettian many-worlds theory of quantum mechanics. 

 In a strongly deterministic world, at the fundamental level, there is no contingency. There is only one way the universe could be, in the sense of nomic possibility. If the actual world is strongly deterministic, then the actual world is the only one that is nomologically possible. That is, what is actual is also nomologically necessary. Any sense of contingency would have to come out at some non-fundamental level.
 
 Let us  contrast that with a more familiar theory---classical mechanics. The only dynamical law $F=ma$ selects a space of possible worlds---the solutions to that equation. It is possible to add to it a macroscopic initial condition law---the Past Hypothesis, which says that the universe started in a low-entropy macrostate.\footnote{For discussions of the Past Hypothesis, \citep{boltzmann2012lectures, feynman2017character, penrose1979singularities, albert2000time, goldstein2001boltzmann, callender2004measures, loewer2012two, loewer2016mentaculus}.} This special macrostate allows the emergence of various asymmetries in time such as the Second Law of Thermodynamics. However, even after adding the Past Hypothesis, there is still an infinity of microscopic initial conditions compatible with the laws. These initial conditions (compatible with both $F=ma$ and the Past Hypothesis) have smaller measure than the original set (whose only constraint is $F=ma$). In this scenario, even though the  laws are deterministic, they do not pick out a unique history of the universe, as there are many possible worlds compatible with the laws. Hence, the laws are not strongly deterministic. 

In a strongly deterministic world,  the laws specify the entire history of the universe without needing any further input. Hence, it is in principle possible to extract all the information about the world from just the laws alone. It is this feature of strong determinism that  makes possible an empirically adequate theory of a cosmic void, in the sense of being informationally complete to recover all the facts. (Is this sense of empirical adequacy strong enough? We return to this question in \S4.) Suppose at the fundamental level we have strongly deterministic laws.  There is only one way the fundamental material ontology could be like---they have to be arranged exactly according to the only way allowed by the laws. Such an arrangement of fundamental matter would be nomologically necessary. Facts about fundamental matter would make true non-fundamental facts about the locations and behaviors of tables and chairs.  

We can go further. The strongly deterministic laws constrain the fundamental ontology \emph{if it exists}. But now suppose that there is no material ontology at the fundamental level. All we have are the laws of physics, from which we can in principle derive all facts about the non-fundamental. Would it make a difference whether the derivation goes through some postulate about fundamental matter? It depends on one's view about ontological dependence: must non-fundamental facts about tables and chairs bottom out in fundamental facts about matter? (We return to this question in \S4.) Suppose we accept the possibility that non-fundamental facts about tables and chairs can bottom out in fundamental facts about laws of nature. That is, all explanation about the non-fundamental can be purely nomic. We arrive at the cosmic void scenario---in a strongly deterministic universe, the physical theory can be empirically adequate without postulating any fundamental material ontology. All there is at the fundamental level are the  laws of nature. 

What is this notion of ``deriving'' non-fundamental facts from laws of nature? I suggest this does not have to differ  from the notion of mathematical-physical derivation that links the macroscopic facts such as the positions of tables to the microscopic facts such as the positions of particles. The derivation combines a variety of techniques, such as mathematical proofs, coarse-graining, approximations, and idealizations. The kind of examples I have in mind are those in statistical mechanics concerning the reduction from the macroscopic phenomena to kinetic theory. But they can also be found in high-energy physics and quantum theory. However, in the more familiar examples in physics, since the laws people use are not strongly deterministic, the laws themselves are insufficient, and the derivation must involve further input about the contingent initial conditions.   

It is worth emphasizing that, since we are assuming a non-Humean theory of lawhood, strongly deterministic laws are compatible with the existence of fundamental material ontology as well as their absence. If we do not postulate fundamental material ontology in a strongly deterministic universe, if the cosmic void scenario is possible, then there can be non-fundamental facts about the material ontology that can coincide with the content of the fundamental material ontology that could be postulated. However, whether or not the facts about such material ontology are fundamental, they make true the same set of non-fundamental facts---the locations of tables and chairs. So it is best to say that the strongly deterministic universe is compatible with two worlds that coincide in all facts except the fundamental material facts. Hence, we need to revise our earlier definition of strong determinism:  
\begin{description}
\item[Strong Determinism*] The laws of nature pick out a unique way to completely specify the fundamental matter. 
\end{description}
However, a strongly deterministic world on this criterion does not need to specify any fundamental matter---it could realize the cosmic void scenario that is devoid of fundamental matter. But if we were to completely specify fundamental matter, then there would  be only one way to do so according to the laws. The qualifier ``completely'' is important here: if we were to only partially specify the fundamental matter, then there could be infinitely many ways to do so. We could, for example, specify only the configuration of matter in spacetime region $R_1$, specify only that in $R_2$, and so on.

The cosmic void scenario has many interesting features. In such a world, even though the fundamental arena (space-time or something analogous to space-time) is completely empty, we can still recover all the non-fundamental facts from fundamental laws alone, relying on purely nomic explanations. That would be surprising if it turns out to be a successful theory. What are some advantages for maintaining the cosmic void scenario? First, it has a fairly parsimonious ontological base. All else being equal, we have reasons to prefer a more parsimonious theory. Of course, not everyone will be convinced that other things are equal here. Perhaps we lose certain explanatory power. We also should be careful where and how to apply  Ockham's Razor for reasons mentioned in \citep{maudlin2007completeness}. We will return to this point in \S4. Second,  the cosmic void scenario also shares some of the advantages of radical versions of ontological nihilism, such as the versions developed in \citep{HawthorneCortensTON}. For example, since there are no material objects at the fundamental level, the difficult metaphysical disputes about identity of material objects over time and mereological composition of material objects simply do not arise, at least for things that exist at the fundamental level. Nevertheless, neither consideration is decisive. Moreover, the cosmic void scenario may cost us  important intuitions or principles that we cherish (see \S4). In the end, it will require a more comprehensive cost / benefit analysis (one that we do not have the space to develop in this paper) to reach a reflective equilibrium about whether the cosmic void scenario should be allowed in logical space.  Let us now turn to a case study of the cosmic void scenario.

\section{A Case Study}

In this section, we provide a concrete example of strong determinism to illustrate the possibility of the cosmic void. The example relies on a non-standard way of combining Everettian many-worlds interpretation of quantum mechanics with the low-entropy initial condition of the universe that is essential for explaining the arrow of time. 

\subsection{Many-Worlds Interpretation of Quantum Mechanics}

Hugh Everett III's many-worlds interpretation (\citeyear{everett1957relative}) was an attempt to address a central problem in the foundation of quantum mechanics---the quantum measurement problem.\footnote{Below is a rough sketch. For a more thorough discussion about the quantum measurement problem, see \citep{bell1990against, albert1994quantum, sep-qt-issues}.} Textbook quantum mechanics suggests that we assign a quantum state, represented by a wave function, to some physical systems of interest, such as a cat in a box. The wave function obeys two dynamics. First, it obeys a linear equation of motion---the Schr\"odinger equation. That equation has the tendency to spread out the wave function into what is called a superposed state, such as the cat-being-alive state superposed with the cat-being-dead state. In that state, it is hard to understand what is going on physically: is the cat alive,  dead, or both? Second, upon ``observation'' or ``measurement,'' the system's wave function will undergo a random and non-linear collapse to a particular state, such as the state that the cat is alive. 

Despite the predictive success, this recipe for making predictions faces  foundational problems. How can the wave function obey two different dynamics? If the system is completely described by the wave function, and if both observers and measurement instruments are physical systems interacting quantum mechanically, they would always obey the  Schr\"odinger equation and never collapse into a definite state (such as the cat-being-alive state) from a superposed state (such as the superposed state of the cat-being-alive and the cat-being-dead). We can put together the principles that result in a contradiction: 
\begin{itemize}
\item[(P1)] The wave function is the complete description of the physical system.
\item[(P2)] The wave function always obeys the Schr\"odinger equation.
\item[(P3)] Every experiment has a unique outcome. 
\end{itemize}

There are three types of solutions to the quantum measurement problem. Each of them rejects one of the assumptions. The first type of solutions rejects (P1) and adds additional variables beyond the wave function to represent other aspects of the system. A prominent theory in this type is Bohmian mechanics, originally proposed by \cite{de1928nouvelle} and \cite{bohm1952suggested} independently. See \citep{sep-qm-bohm} for an overview. On that theory, there are point particles with precise locations in addition to the wave function. The cat was made out of particles guided by the wave function. The particles are always in a determinate macrostate---either the cat is alive or the cat is dead. The wave function plays two roles: it guides the motion of particles and it provides a probability distribution over their precise locations. 

The second type of solutions rejects (P2) and replaces the Schr\"odinger equation by a spontaneous collapse dynamics. A theory in this type is the GRW theory, proposed by Ghirardi, Rimini, and Weber (\citeyear{ghirardi1986unified}). See \citep{sep-qm-collapse} for an overview.   The GRW theory  replaces the measurement-triggered collapse dynamics with something much less anthropocentric---an objective and spontaneous collapse mechanism that collapses the wave function with  a fixed probability per unit time per unit particle. Before we open the box to measure the cat, its wave function has already collapsed into one of the two states: either alive or dead. 

The third type of solutions rejects (P3) and embraces the non-uniqueness that comes with it. This was proposed by Everett  and also known as the many-worlds interpretation of quantum mechanics. See \citep{sep-qm-manyworlds} for an overview.  On this theory, the wave function is the complete description of the physical system. It always obeys the Schr\"odinger equation. But typically experiments lead to multiple definite outcomes all at once. When we go and measure the cat, we observe that it is in a definite state, say, the cat is alive. But appearances can be misleading. Since nothing has collapsed the wave function and there are no additional independent variables, the part of the wave function in which the cat is dead  still exists and is equally real as the part in which the cat is alive. What is going on? On this interpretation, there are at least two ``parallel worlds'' after the experiment that correspond to the two outcomes. Both are real---they are known as two branches of the wave function. Usually after the experiment, the different branches no longer interact much with each other. Whether the parallel worlds come from a splitting of a single world into two or are merely emergent descriptions is a matter of debate. As with \cite{wallace2012emergent}, I think it is much more plausible to interpret the parallel worlds from the emergence perspective---they are not fundamental properties of the Everettian universe. 

\emph{Unlike} the GRW theory and \emph{like} Bohmian mechanics, the many-worlds interpretation is deterministic: given a complete specification of the state of the universe at one time, all the past and all the future is completely fixed by the law of motion. In this case, the wave function of the universe at one time, by the Schr\"odinger equation, completely fixes the wave function at all other times.  But the many-worlds interpretation differs from Bohmian mechanics in that the latter has a single world while the former entails a multiplicity of (emergent) worlds. The reason is that the wave function, which is ontologically complete in the Everettian picture, introduces a multiplicity that is resolved in the Bohmian picture by the additional postulate of point particles with precise locations. 

What is the fundamental material ontology on the many-worlds interpretation? On the most flat-footed way of thinking about it, the fundamental material ontology consists in the quantum state, represented by the wave function. However, what the wave function represents in the physical world is a controversial matter.  Realism about the wave function is  compatible with many different  views about the underlying nature of the physical reality represented by the wave function. See \citep{chen2019realism} for an overview. There are two interpretations that will be relevant below. First, we can use the wave function of the universe to define a matter distribution on physical space-time. This is known as Sm and proposed first in \citep{allori2010many}. The basic idea is that we can use the wave function to define a function $m(x,t)$ whose domain is physical space-time and whose range is the set of real numbers. This function can then represent a physical field called the matter-density field in space-time. The higher the ``sum'' of values of the  $m(x,t)$ function in some region $R$, the more stuff there is in $R$. The locations of tables and chairs at different times can be read off from the history of this function $m(x,t)$.    Second, we can follow \cite{wallace2010quantum} and postulate local quantum states for sub-regions of the universe. Given a partition of the universe into collections of sub-regions, we can use the wave function of the universe to define sub-region states. (More technically, we can define reduced density matrices for some local region $R$ by tracing out the degrees of freedom of the environment $E$ in the universal wave function.) Because of the holism inherent in quantum entanglement, merely defining the reduced states for each sub-region is not sufficient to describe the history of the universe. In general, even after we postulate the reduced state of $R_1$ and the state of $R_2$, we will still need to postulate a state of $R_1  \bigcup R_2$, as its state is in general not a logical sum of the states of its parts.  

We will focus on the many-worlds interpretation of quantum mechanics here because it has the potential to develop into a strongly deterministic theory that realizes the possibility of a cosmic void. 

\subsection{The Everettian Wentaculus}

The many-worlds interpretation is already deterministic. To make it strongly deterministic, it suffices to pick a unique microscopic initial condition of the universe. One idea is to pick a particular wave function and call it the actual and nomologically necessary initial wave function of the universe. But that would not work. Typical wave functions of the universe contain too much information that they will in general be extremely complicated. It is unlikely that a simple law can pick out any of those complicated wave functions. 

In this section, we  outline a new version of many-worlds interpretation whose laws  are strongly deterministic \emph{and} simple. The laws will be: the von Neuman equation that generalizes the Schr\"odinger equation,  the Initial Project Hypothesis that replaces the Past Hypothesis, and possibly an additional  matter-density equation. 

To appreciate the new version of many worlds interpretation of quantum mechanics, we need to introduce some background about the arrows of time. We see the manifestation of arrows of time everywhere. Many processes in nature typically happen in only one direction: they are time-asymmetric. In room temperature, the ice cubes in the past  are always larger than the ice cubes in the future; bananas are more ripe in the future than in the past; people are more wrinkled in the future than in the past. However, the fundamental dynamical laws of physics are (essentially) symmetric between the past and the future. Take for example the conjunction of Newton's equation of motion $F=ma$ and law for universal gravitation $F=Gm_1m_2/r^2$, whatever can happen forward can  as easily happen backward. The same is more or less true for the Schr\"odinger equation and special and general relativity. So the explanation for the arrows of time plausibly come from elsewhere. 

One prominent explanation, first proposed by \cite{boltzmann2012lectures} (published first in 1899), invokes what is now called the Past Hypothesis---the universe in the beginning had very low entropy, where entropy is a measure of disorder. Given a low-entropy initial macrostate,  the universe will typically develop in such a way to exemplify arrows of time in its typical subsystems. That is, typically, typical subsystems of the universe are entropic. Modern developments of Boltzmann's ideas in the foundations of statistical mechanics further substantiate this idea. The Past Hypothesis is still compatible with  infinitely many initial microstates. It is reasonable to assume that some of them will lead to anti-entropic behaviors  (e.g. ice cubes will become larger in room temperature). To solve this problem, it is essential to have the Statistical Postulate according to which these  anti-entropic initial states have relatively small measure---they are atypical. \cite{albert2015after} and \cite{loewer2016mentaculus} call the conjunction of the dynamical law, the Past Hypothesis, and the Statistical Postulate \emph{the Mentaculus}. 

Albert and Loewer's Mentaculus theory was developed in the classical domain. In the quantum case, we can postulate something similar to the Mentaculus. In the quantum case, the Past Hypothesis is now a constraint on initial wave functions of the universe---all of the nomologically possible ones started in a low-entropy macrostate $\mathscr{H}_{PH}$, which is a extremely small subset of all available wave functions. The wave functions inside the macrostate are macroscopically similar in entropy, temperature, volume, and pressure. Inside this macrostate, every wave function is equally likely as any other. (To be more rigorous, we say that there is a uniform probability distribution on wave functions compatible with $\mathscr{H}_{PH}$ with respect to the normalized surface area measure on the unit sphere of the Hilbert space subspace picked out by $\mathscr{H}_{PH}$.) Given the Schr\"odinger equation, the wave function of the universe will rotate continuously inside the full energy shell of the Hilbert space. That will describe (deterministically) the history of the universe for all times. 

In the quantum case, however, another framework is possible to combine the low-entropy initial macrostate with the quantum dynamics. The alternative framework, which I call \emph{the Wentaculus}, has surprising consequences, including the realization of strong determinism.  I introduce this framework in \citep{chen2018IPH} and develop it in other work such as \citep{chen2019quantum1, chen2018valia, chen2018NV}.   Instead of postulating an initial macrostate compatible with many initial microstates, we can use the initial macrostate  $\mathscr{H}_{PH}$ to select a unique and natural microstate---the normalized projection onto that subspace. This is represented by a mixed state density matrix $W$. This statement is captured in what I call the \emph{Initial Projection Hypothesis} (IPH), a candidate new law of nature that replaces the Past Hypothesis in the Mentaculus. (IPH is as simple as the Past Hypothesis.) Given that there is only one initial microstate possible under IPH, we no longer need to impose the Statistical Postulate to neglect abnormal initial states. Moreover, we can replace the wave function dynamics by the corresponding density matrix dynamics, which is still deterministic in the case of Everettian theory.  In this  version of Everett, then, the dynamics is deterministic and there is only one microscopic initial condition compatible with the laws. Thus, this version of Everett is strongly deterministic: given the laws, there is only one microscopic history of the universe that is possible. That makes the actual history of the Everettian universe nomologically necessary.\footnote{More precisely, we can write down the postulates of this version of Everett. The Initial Projection Hypothesis is as follows:
\begin{equation}
\hat{W}_{IPH} (t_0)  = \frac{I_{PH}}{dim \mathscr{H}_{PH}}
\end{equation}
where $I_{PH}$ is the projection operator onto the subspace $ \mathscr{H}_{PH}$ and ``dim'' counts the dimension of that subspace. 

 The Von Neumann Equation is as follows:
\begin{equation}
i \hbar \frac{\partial \hat{W}}{\partial t} = [\hat{H},  \hat{W}]
\end{equation}
where the commutator bracket denotes the linear evolution generalized from the Schr\"odinger equation. 
}  

We arrived at a strongly deterministic version of  \emph{the Wentaculus} (``W'' for the initial density matrix). Not all versions of the Wentaculus are strongly deterministic. The Bohmian version is not strongly deterministic because there are additional independent variables representing possible initial locations of point particles. The GRW version is not strongly deterministic because the dynamics is stochastic. 

The Everettian Wentaculus theory we outlined above admits two interpretations: one that is compatible with a fundamental material ontology and the other compatible with the cosmic void scenario. For the first one, we can postulate that at the fundamental layer of reality there is matter: the quantum state, the reduced states, or the matter-density ontology represented by $m(x,t)$.\footnote{For the matter-density ontology, we can postulate the following law---the Matter Density Equation: 

\begin{equation}\label{mxt}
m(x,t) = \text{tr} (M(x) W(t))
\end{equation}
where $x$ is a physical space variable, $M(x) = \sum_i m_i \delta (Q_i - x)$ is the mass-density operator, which is defined via the position operator $Q_i \psi (q_1, q_2, ... q_n)= q_i \psi (q_1, q_2, ... q_n) $.} 
The non-fundamental facts  and the manifest image are recovered by using facts about fundamental material ontology as well as the laws.  

However, it also supports a second interpretation that is compatible with the cosmic void scenario. Since the laws are sufficient to determine the actual microstate of the universe at all times, following the arguments in \S2, we no longer need to postulate facts about fundamental material ontology to explain the non-fundamental facts. We can bypass the fundamental ontology and to  derive facts about the locations of tables and chairs directly from the laws---in this case the von Neumann equation that determines how density matrix evolves and the Initial Projection Hypothesis that fixes the initial density matrix. However, the language we use  here can be misleading. The equation and the hypothesis seem to assume the existence of some material object represented by the density matrix. But it does not have to exist at the fundamental level. All we need is the information stored in the laws of nature. (What are the laws about if they are not about some fundamental objects changing in time? We return to this question in \S4.)

The Everettian cosmic void scenario, if it can be made empirically adequate,  has the following features:
\begin{enumerate}
	\item The fundamental laws are strongly deterministic. There is no contingency at the fundamental level.
	\item There is no material ontology at the fundamental level (no quantum state, reduced states, or matter density field). 
	\item At the non-fundamental level, there is an emergent multiplicity of worlds that can be characterized by non-fundamental material ontologies such as the quantum state, reduced states, or a matter density field. 
	\item The non-fundamental facts can be derived from   facts about the fundamental laws of nature. 
	\item At the non-fundamental level, contingency can reappear as descriptions of the non-fundamental worlds. 
\end{enumerate}

The Everettian Wentaculus provides a concrete example of the possibility of the cosmic void. In such a world, all there is at the fundamental level are the strongly-deterministic laws. The non-fundamental facts are derived from them.\footnote{The status of space-time in this theory is an interesting issue. A defender of the Everettian cosmic void scenario may entertain the possibility that space-time is not fundamental and is also emergent from the strongly deterministic laws. She may use ideas from the research program initiated in \citep{carroll2018mad}.  However, this move will be in tension with the arguments discussed in \citep{north2017new}. }

\section{Challenges}

Can the Everettian cosmic void scenario be made empirically adequate? That question depends on a further question: can the Everettian picture be made empirically adequate? Mover, even if the standard Everettian picture can be empirically adequate, the Everettian cosmic void may suffer from additional metaphysical and semantic challenges.  We discuss some of them in this section. 

\subsection{The Semantic Challenge}
We start with a semantic challenge: What are the fundamental laws about if there is no material ontology at the fundamental level? What do the theoretical terms refer to? 

On the standard picture, the fundamental laws are about the fundamental material ontology. The terms in the fundamental laws refer to properties of  fundamental matter (e.g. the mass of point particles). Since there is no material ontology in the cosmic void scenario, this way of thinking about \textit{aboutness} and reference need to be revised.

One possible defense of the cosmic void scenario is to invoke certain non-fundamental objects to make sense of the laws.  The Initial Projection Hypothesis, for example, refers to an initial density matrix. But what is the meaning of the initial density matrix if there is nothing material at the fundamental level of which it is about? One possibility is to appeal to the effects it has on the \emph{non-fundamental} worlds and objects. To understand the meaning of the Initial Projection Hypothesis, it may suffice to understand it in terms of the behaviors of the tables and chairs, which are derived from the hypothesis. After all, we make sense of fundamental laws by their effects on observable objects, such as the behaviors of pointers and measurement instruments. How their behaviors are connected to more fundamental facts requires theoretical postulates. Our epistemic access to the fundamental material ontology is rather limited anyways.\footnote{Thinking about the metaphysics of fundamentality, \cite{bernsteincould} suggests that we should take seriously the possibility that the middle level inhabited by medium-sized dry goods such as tables and chairs is the most fundamental level of reality. Middle level fundamentalism provides an alternative to the usual debate between microphysical-fundamentalism and cosmos-fundamentalism. This is not the metaphysical picture I favor (and neither does Bernstein endorse middle level fundamentalism). However, if that view is possible, then presumably the fundamental laws in such a world  are primarily about the behaviors of tables and chairs. Hence, it should also be possible that we can make sense of  the fundamental laws in the cosmic void scenario by looking at their effects on the non-fundamental objects. } 

To be sure, some might object that fundamental laws cannot be about anything other than fundamental things or properties.\footnote{For example, see discussions in \citep{LewisNWTU, SiderWBW}.} If one holds this view and thinks this view is justified, one would dismiss the possibility of the cosmic void scenario. However, one might think  it is \textit{better} if the laws refer only to fundamental objects and their properties but remain  agnostic whether we are justified in imposing such a condition on all physical theories. One can maintain that the principle is  a desirable theoretical virtue to be balanced with other considerations. For example,  \cite{hicksschaffer} suggest that this principle is not always true. However, I think that it is a significant cost  if we have to give up this principle to entertain the cosmic void scenario.

\subsection{The Metaphysical Challenge}
Next, we have a metaphysical challenge: How can non-fundamental facts about tables and chairs depend on facts about non-matter (laws)? 

On the standard picture, the non-fundamental facts are ultimately explained (at least in part) by fundamental facts about fundamental matter. For example, a typical reductionist would like to reduce facts about tables and chairs into facts about the arrangements of particles and configurations of fields. In the cosmic void scenario, the fundamental facts do not include such facts. So the old way of thinking about ontological dependence needs to be revised if the non-standard picture is to succeed. 

A defender of the cosmic void scenario may point to an ambiguity about ``fundamental facts about fundamental matter.'' At the fundamental level, there may not be any  facts about matter. However, at some non-fundamental level, there can be facts about matter that is as fundamental as it gets in the cosmic void picture. For example, in the Everettian cosmic void scenario, the bottom level of reality consists in fundamental laws: the von Neumann equation, the Initial Projection Hypothesis, and the matter-density equation. From those equations, we can derive the values of the matter-density field, which can exist non-fundamentally but at a level that is more fundamental than any other material objects. Tables and chairs would emerge at a much higher level from the matter-density field. Therefore, there can be a vestige remaining about the intuition that non-fundamental facts of tables and chairs should bottom out in terms of fundamental matter: it is just that the most fundamental facts about matter will be explained by even more fundamental facts about laws of nature. 

Another metaphysical challenge to the cosmic void scenario is related to the issue of completeness in  the foundations of quantum mechanics. Is the wave function all there is or is there something else in the material ontology? In the framework of GRW theory, several answers are available \citep{allori2008common}. First, there is the bare GRW theory (GRW0) where the only thing in the material ontology is the wave function. Second, there is the GRW theory with a matter density ontology (GRWm) that is similar to the matter-density version of many-worlds interpretation (Sm) we discussed in \S3.1. The definition of the matter density is the same in both theories. However,  they differ in the dynamics of the wave function. They also have different interpretations of the nature of probability. Third, there is the GRW theory with a flash ontology (GRWf), where the matter-density ontology is replaced by discrete events in space-time. 

Both GRWf and GRWm postulate something fundamental and material in physical space-time. (The flash ontology and the matter-density ontology are called the \emph{primitive ontology} in the literature.) The configuration of matter in space-time are derived from the universal wave function. So in principle we just need the information about the wave function to derive the information about everything else. \cite{maudlin2007completeness} suggests that even though the GRW wave function is \emph{informationally complete} in this sense, it should not be thought of as \emph{ontologically complete} in the sense that all there is in the theory is the wave function. After all, which material ontology is privileged in the GRW theory? Is it the flash ontology or the matter-density ontology? They are two different ways to define the material ontology in space-time. A defender of GRW0 may respond that both ontologies can be taken as derivative physical structure that we can use to make precise the connection between theory and evidence. However, whether the response is acceptable may depend on one's  view about primitive ontology. 

We can ask a similar question about the cosmic void scenario. Even though we can derive everything from the strongly deterministic laws, it only shows that the laws are informationally complete, and it does not follow that they are ontologically complete. It requires a further assumption and perhaps additional justification to postulate that the laws are all there is. 

At this point, a defender of the cosmic void scenario may appeal to Ockham's razor: it is more parsimonious to postulate just the strongly deterministic laws than to postulate both the laws plus some material ontology. This is different from the application of the razor in a merely deterministic universe: since we can derive all facts about the past and the future from the laws and the complete state of the universe at some time, perhaps we can get rid of all the other times \citep[p.~3153]{maudlin2007completeness}. We can care about minimizing the kinds of things without caring so much about minimizing the number of things in the same category. The application of Ockham's razor in the strongly deterministic universe gets rid of an entire category of things: fundamental material ontology. The application of the razor in a merely deterministic universe only gets rid of the vast number of times saving just one. 

There is a strong intuition that no theory can be entirely satisfactory  unless it postulates some kind of ordinary material ontology. However, if one is antecedently sympathetic to a \emph{wave function monist} interpretation of GRW and accepts GRW0 as a satisfactory physical theory, perhaps she should be more open to reject the intuition. On GRW0, the quantum state (represented by a wave function) is the only thing that exists. But it is nothing like the ordinary material ontology of particles and fields. It is defined on a vastly high dimensional space, and the recovery of ordinary facts about tables and chairs is through some abstract mathematical derivations that look nothing like the ordinary ``causal'' explanations in space-time. Someone who accepts GRW0 will be happy to accept that ordinary facts about tables and chairs are just emergent patterns in the high-dimensional wave function. But the cosmic void scenario also locates the non-fundamental facts in some unfamiliar fundamental facts---facts about fundamental laws of nature. If one is happy to accept the emergence of non-fundamental facts from a radical material ontology, then one should be more open to accept the emergence of non-fundamental facts from strongly deterministic laws. 

%Nevertheless, no one is obliged to interpret a strongly deterministic theory in the cosmic void fashion and no one has to use the razor in the way discussed above. What we want to point out in this paper is merely that it is a possible interpretation with interesting metaphysical features. 

\subsection{The Empirical Challenge}
Finally, we have an empirical challenge: How can the Everettian Wentaculus theory be empirically adequate when every possible outcome of experiment is realized? How can  the Born rule of probability make sense in such a world? 

This is a familiar challenge to many-worlds interpretations of quantum mechanics. It applies to both the Everettian Mentaculus  and the Everettian Wentaculus (as well as the cosmic void scenario). It is also related to the emergence of contingency in a strongly deterministic universe. 
 All versions of the many-worlds interpretation face a general problem of probability:  if all outcomes of the experiments actually occur, how do we make sense of non-trivial Born rule probabilities? For example, in the case of Schr\"odinger's cat, we can set up the experiment such that quantum mechanics says the probability that the cat will be alive is 0.3 and the probability that the cat will be dead is 0.7. But if both outcomes are realized the cat is alive in one branch and the cat is dead in another, what to make of such probabilities? Defenders of Everett, such as \cite{deutsch1999quantum} and \cite{wallace2012emergent}, appeal to Savage-style decision theory. There is also a line of defense developed by  \cite{sebens2016self} that appeal to rational norms governing self-locating probabilities.\footnote{Everett himself appealed to arguments about long run frequencies and typicality of branches. See \citep{barrett2017typical}. } 

But these defenses of Everett make two kinds of assumptions. First, contrary to the Wentaculus, the quantum state of the universe is assumed to be pure, represented by a wave function. A technical problem for the Everettian Wentaculus is how to adapt their arguments in the case when the universe is in a fundamental  mixed state, represented by a density matrix. The technical challenge, in my opinion, is not difficult to overcome but there is work to be done. Second, they also postulate some principles of rationality in order to derive the Born rule. In the case of \cite{wallace2012emergent}, the Savage-Wallace axioms of preferences are far from being rationally obligatory. In the case of \cite{sebens2016self}, the epistemic separability principle seems to be open to counter-examples. So the hard problem is to make a strong case to justify these assumptions in the proof of the Born rule. I am pessimistic about the prospects of this project, but perhaps the difficulties can be overcome. If the technical problem of extension of the argument to mixed states can be solved, then the Everettian Wentaculus fares no worse than the standard Everettian theories in the literature, at least with respect to the problem of probability. 

If this project of justifying Born-rule probability in the Everettian framework can succeed, it can also help answer the question: how can contingency emerge in a strongly deterministic Everettian universe like the Everettian Wentaculus? The answer lies in the emergence of the branching structure with a multiplicity of different macroscopic histories (worlds) and the interpretation of branch weight as probability. We leave this project to future work.

\section{Conclusion}

I have introduced a non-standard picture of the physical universe called the cosmic void. I have discussed the general possibility as well as a concrete example in the Everettian Wentaculus theory. The possibility of the cosmic void is tightly connected to the possibility of strong determinism. In the cosmic void scenario, at the most fundamental level of reality, there are only fundamental laws of nature and no material ontology. Facts about matter can emerge at a non-fundamental level: they can be derived from the laws if they are strongly deterministic. The possibility is certainly unfamiliar and faces  difficult challenges, but it deserves careful and critical scrutiny. I hope the picture outlined above can interest other people and lead to further explorations of the topic. 

There is an interesting connection between the cosmic void scenario and the ontologist nihilist's picture of a fundamental ontology without any objects. According to ontological nihilism, there are no objects at the fundamental level. This view is discussed in, for example, \citep{HawthorneCortensTON} and \citep{TurnerON}. \cite{DasguptaI} develops a related view where facts involving individual objects are paraphrased away. \cite{Maxwell2019} also develops a closely related view according to which objects do not play an indispensable role in metaphysics. In the nihilist / anti-object framework, ordinary sentences involving objects are to be paraphrased in terms of non-objectual  language that does not ontologically commit us to the existence of objects. 

The cosmic void scenario is compatible with that picture, but the two approaches are quite different. The heavy lifting in the cosmic void scenario is done not by revising logic or semantics, but by using a particular kind of physical theory to derive non-fundamental facts about matter from fundamental laws alone. Hence, the cosmic void possibility is at the same time more radical and more modest. It is more radical in that we do not postulate even fundamental properties, general facts, bare facts, or structural facts \citep{ladyman2007every} in the fundamental ontology. It is more modest in that the strategy is not supposed to work in general, but only in some special physical theories where strong determinism holds. Hence, we have two potential strategies for getting rid of fundamental material ontology: the metaphysical strategy and the scientific strategy. Both are difficult positions to defend. On the metaphysical side, we need to show that the relevant kind of semantic and logical maneuver results in a satisfactory paraphrase of the original object-language. On the scientific side, we need to show that (1) the actual laws are strongly deterministic and (2) a strongly deterministic cosmic void scenario can be empirically adequate. 

Can they be made to work? I am not sure. Even if neither strategy works, we can learn something from these examples: they may give us additional reasons why material objects are crucial for formulating successful physical theories and writing the book of the world.

%------------------------------------------------

\section*{Acknowledgement}

I would like to thank  the editors of this volume---Sara Bernstein and Tyron Goldschmidt---for useful written comments on an earlier draft of this paper. I am also grateful for helpful discussions with David Chalmers, Sam Elgin, Bixin Guo, Mario Hubert, Joseph Martinez, Mark Maxwell, Daniel Rubio, Ayoob Shahmoradi, J. Robert G. Williams, and the audience at the 2020 Central APA Symposium ``Metaphysics Without Ontology.'' 

%----------------------------------------------------------------------------------------
%	BIBLIOGRAPHY
%----------------------------------------------------------------------------------------

\bibliography{test}

%----------------------------------------------------------------------------------------

\end{document}